\documentclass[%
 preprint, 
 amsmath,amssymb,
 aps, physrev,
]{revtex4-2}
\usepackage{graphicx}
\usepackage{dcolumn}
\usepackage{bm}
\usepackage{xcolor}
\usepackage{hyperref,comment}

\begin{document}


\title{\textbf{
Towards synergy between theory and experiments: First-passage processes under resetting with stochastic return protocols} 
}%

\author{Ashutosh Dubey}
\email{ashutoshrd@imsc.res.in}
\author{Arnab Pal}%
\email{arnabpal@imsc.res.in}
\affiliation{%
 The Institute of Mathematical Sciences, CIT Campus, Taramani, Chennai 600113, India \& Homi Bhabha National Institute, Training School Complex, Anushakti Nagar, Mumbai 400094, India 
}%

\begin{abstract}
Stochastic resetting with stochastic return (SR) has recently emerged as an effective and practically relevant strategy for controlling first-passage phenomena and optimizing search processes by significantly reducing completion times in a wide class of stochastic systems. While most theoretical studies assume instantaneous return (IR), in realistic systems the return to the resetting position typically occurs over a finite time and follows its own dynamics. We consider a one-dimensional diffusing particle confined in a harmonic trap, with the trap stiffness taking two distinct values $k_1$ in the search phase and $k_2$ in the return phase. The diffusing particle initially undergoes the search phase, and if it does not reach the target, it is reset at random times occurring at rate $r$, after which it enters the return phase. The particle remains in this return phase until it either reaches the resetting position (the origin) or the absorbing target located at $L>0$. Notably, the model permits target detection throughout the return phase, distinguishing it from IR scenarios. We analytically derive the mean first-passage time for our model and determine parameter regions where SR enhances or suppresses search efficiency over the underlying dynamics, constructing the corresponding phase diagram using the coefficient of variation criterion. We further identify regimes in which SR is more efficient than IR using the stochastic return criterion. This behavior is tested across different parameter sets taken from experimental studies by Goerlich \textit{et al.} \cite{goerlich2025taming}, Biroli \textit{et al.}\cite{jkn8-h939}, and Besga \textit{et al.} \cite{besga2020optimal,besga2021dynamical}. We map the parameter values reported in existing experimental studies onto our theoretical phase diagram, thereby identifying the regimes in which SR outperforms IR. In particular, we find that the most relevant region of the phase space, where SR is most beneficial corresponds to larger values of the potential strength in the search phase and lower values of the potential strength in the return phase.  Our results provide experimentally testable predictions and bridge the gap between theoretical resetting models and optical trap mediated finite-time return dynamics.
\end{abstract}

\maketitle


\section{Introduction}

Stochastic resetting has emerged as a versatile framework for controlling nonequilibrium stochastic dynamics and optimizing first-passage processes. By interrupting a system's natural evolution and restarting it from a prescribed state, resetting can profoundly alter its temporal properties, giving rise to nonequilibrium steady states \cite{Evans2011,majumdar2015dynamical,evans2018run,magoni2020ising,pal2015diffusion,basu2019symmetric,gupta2020work,kumar2023universal,di2023current,mendez2016characterization} and substantially modifying search efficiency. Its applications span a broad range of problems, including diffusive search and transport \cite{campos2015phase,chechkin2018random,de2022optimal,chen2022first,ahmad2019first,biroli2026first,garcia2023optimal,ray2020space,ray2019peclet,evans2025stochastic,reuveni2016optimal,pal2022inspection,dubey2023first,pal2017first,pal2024random,thakur2026resetting,pal2025universal,sandev2022heterogeneous,linn2026dynamic,jain2023fick,pal2024channel,barman2025optimizing,pal2023thermodynamic}, animal foraging \cite{pal2020search,kusmierz2014first}, molecular motor transport and facilitated diffusion \cite{bressloff2020directed,rajoria2026broad,loverdo2009quantifying}, and biochemical reaction kinetics \cite{reuveni2014role}. Beyond search, resetting has also proved useful in describing and controlling other stochastic processes, including queueing systems \cite{roy2024queues,bonomo2022mitigating}, and extreme-event statistics \cite{montanari2002optimizing}. Related restart mechanisms have also been explored in computational search algorithms \cite{hamlin2019geometry,luby1993optimal} and in economic and financial settings \cite{jolakoski2023first}. These diverse applications have motivated sustained interest in understanding how resetting protocols influence the dynamics, efficiency, and statistical properties of stochastic systems. For a detailed overview of the field, we refer the reader to the following reviews \cite{evans2020stochastic,pal2024random,gupta2022stochastic,pal2022inspection,keidar2026stochastic}.

In physical systems, resetting to a prescribed position generally requires a finite amount of time, during which the particle returns to the reset position according to a prescribed or dynamically evolving process. This motivates extensions of the idealized instantaneous return (IR) framework to finite-time return protocols. In deterministic return protocols, the return dynamics is prescribed, for example, through motion at a constant velocity or through an externally imposed protocol \cite{pal2019invariants,pal2019time,pal2020search,bodrova2020resetting,radice2021one,stanislavsky2022subdiffusive,bressloff2020search,bressloff2020target,maso2019transport,radice2022diffusion,mercado2020intermittent}. Alternatively, the return dynamics can be stochastic, with the return time emerging from the particle's own dynamics, such as drift in a fluctuating environment or relaxation in a confining potential \cite{gupta2021resetting,gupta2021stochastic,santra2021brownian,gupta2022work,biroli2024dynamically,frydel2025search}. Such stochastic-return protocols have been investigated in the context of steady-state properties and other aspects of the dynamics. Their implications for first-passage processes have received attention more recently \cite{biswas2024search,biswas2025drift}. In earlier studies, the target was assumed to be inaccessible during the return phase, so absorption could occur only during the search phase. More recently, Refs. \cite{biswas2024search,biswas2025drift} introduced stochastic-return protocols in which the target remains accessible during return, allowing the particle to reach the target before completing its return to the reset position. This distinction introduces an additional pathway to first passage and motivates further investigation of search processes in which the return dynamics is explicitly incorporated into the resetting protocol.

Alongside these theoretical developments, stochastic resetting has also been realized experimentally, providing opportunities to investigate resetting protocols in controlled physical systems. An early experimental implementation by Tal-Friedman \textit{et al.} \cite{tal2020experimental} employed a diffusing particle that searched for an exponentially distributed random duration before an optical trap was activated to return it to a prescribed location, either at constant velocity or within a fixed return time. Subsequent experiments explored stochastic return protocols in which the particle undergoes a random search phase followed by relaxation in a confining potential toward an equilibrium distribution \cite{besga2020optimal,besga2021dynamical,goerlich2025taming,jkn8-h939,ginot2026experimental,goerlich2026time}. Although the return process in these experiments is finite-time and stochastic, the recorded trajectories were analyzed by removing the relaxation segments, thereby effectively mapping the dynamics onto an instantaneous-resetting description. More recently, experimental studies have extended instantaneous stochastic resetting to self-propelled particles, opening new avenues for exploring resetting in active systems \cite{paramanick2024uncovering,kundu2025emulating}.

In this work, we investigate first-passage dynamics under stochastic resetting with stochastic return (SR), considering a diffusing particle whose dynamics is governed by an Ornstein--Uhlenbeck (OU) process during both the search and return phases. Resetting events occur at exponentially distributed random times, initiating a return phase in which the particle relaxes toward the reset position under a harmonic potential. The two phases are characterized by different trap stiffnesses, thereby realizing a switching-trap protocol. Unlike conventional IR, here the return phase has a finite, stochastic duration, and the target remains accessible throughout this phase. Consequently, the particle can reach the target either during the search phase or while returning to the reset position, creating an additional pathway to first passage. We analytically derive the mean first-passage time (MFPT) and compare the efficiency of SR with that of IR, identifying parameter regimes in which finite-time return can enhance search performance. Motivated in part by experiments involving switching-trap dynamics, in which relaxation segments have been excluded from recorded trajectories to construct an effective instantaneous-resetting description, we retain the full search-and-return dynamics and examine the consequences of explicitly accounting for target absorption during return. By connecting our theoretical predictions with parameter values reported in existing experiments, we further identify regimes that could motivate future experimental tests of the benefits of SR.


In this work, we build on the SR framework developed by Biswas \textit{et al.} \cite{biswas2024search,biswas2025drift} to investigate the first-passage properties of a particle evolving under switching harmonic trap potentials. In Sec. \ref{sec2}, we analytically derive the MFPT for the proposed model. In Sec. \ref{sec3}, we employ the coefficient-of-variation (CV) criterion to construct phase diagrams identifying parameter regimes in which SR enhances or suppresses search efficiency. We further use the SR criterion to compare SR with IR and identify regimes in which finite-time return is advantageous. In Sec. \ref{sec4}, we connect our theoretical predictions with existing experiments by mapping parameter values reported in the literature onto the phase-space diagram and examining whether they fall within the regimes favoring SR or IR. Finally, in Sec. \ref{sec5}, we summarize our findings and discuss directions for future research.

\section{\label{sec2} Model System: Switching Harmonic Potential}
\begin{figure}
    \centering
    \includegraphics[scale=0.5]{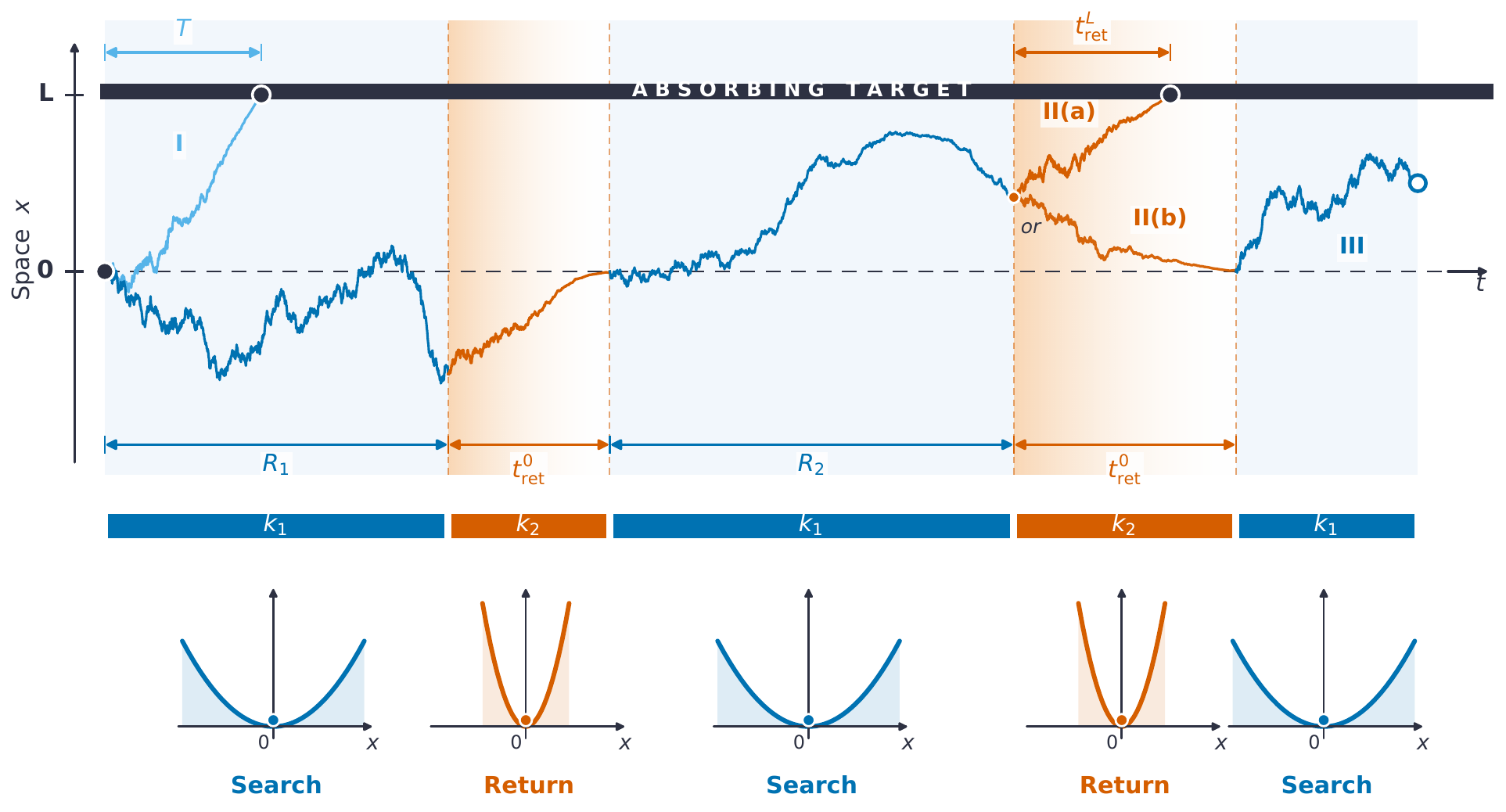}    
    \caption{Schematic illustration of a search process with stochastic resetting with SR facilitated by switching potential dynamics. The initial and resetting positions of the searcher are at the origin. The target is located at a distance $L$ from the origin. The parameters $k_1$ and $k_2$ denote the potential strengths in the search and return phases, respectively, with the potential minimum at the origin. The blue trajectories correspond to the search phase, while the orange trajectories represent the return phase. We illustrate three possible scenarios: (i) the searcher reaches the target directly without resetting in time $T$ shown by blue trajectory (I); (ii) during the return phase, since the searcher is not blind, it may either reach the absorbing target in time $t^L_{ret}$ shown by II(a) or return to the resetting location (here, the origin) in time $t^o_{ret}$ shown by II(b); and (iii) upon returning to the origin in the return phase, the searcher resumes the search process shown by (III).}
    \label{schematic}
\end{figure}
We investigate a one-dimensional search process with stochastic resetting and SR, in which both the search and return phases are governed by OU dynamics. Without loss of generality, we assume that both the initial and resetting positions are located at the origin. During the search phase, the particle starts from the initial position $\bar{x}_0$ and evolves under a confining harmonic potential of stiffness $\bar{k}_1$ with diffusion constant $D$ and friction coefficient $\gamma$, diffusing toward an absorbing target located at a fixed distance $L$. If the particle hits the target in the search phase, the first-passage process terminates. Otherwise, a resetting event initiates the return phase. Resetting events occur at random times drawn from an exponential distribution $f_R(\bar{t}) = \bar{r}e^{-\bar{r}\bar{t}}$, where $\bar{r}$  denotes the resetting rate. At the resetting time, the trap switches to $k_2$, keeping the values of the other parameters the same as those of the search phase. In this phase also, we again have two possible outcomes: since the dynamics is stochastic, if the particle reaches the target before resetting to the origin, at a time denoted by, $t_{ret}^L$ and the process then ends. The second possibility in the return phase is that the particle first returns to the origin, at a time denoted by  $t_{ret}^O$. The return phase then ends, and a new search phase starts from the reset location. The process continues until the particle reaches an absorbing boundary (target) located at position $L>0$, as shown in schematic Fig. \ref{schematic}. The dynamics of the particle are governed by the overdamped Langevin process. We introduce a rescaled, dimensionless length scale $x = \bar{x}/L$ and time scale $t = \bar{t} D / L^2$, and write the Langevin equation in terms of these rescaled variables. The resulting dimensionless form of the Langevin equation is given by
\begin{align}
	\frac{dx}{dt}=
	\begin{cases}
		   &-k_1 x +\sqrt{2}~\eta(t) , ~~~~~~~~\text{search phase }, \\
		  &  -k_2 x +\sqrt{2}~\eta(t),  ~~~~~~~~\text{return phase }.
	\end{cases}
	\label{eom}
\end{align}
where the dimensionless potential traps in search and return phases are related with the dimensional variable by $k_1 = \frac{L^2}{\gamma D}\bar{k}_1$, $k_2 = \frac{L^2}{\gamma D}\bar{k}_2$ respectively, and $\eta(t)$ is Gaussian white noise with zero mean and unit variance. So, in the rescaled description, the problem reduces to a search process that starts and resets at the same position $x_0=\bar{x}_0/L$, continuing until the particle is absorbed at $x=1$ within the domain $(-\infty,1]$. We next establish a general framework for computing the MFPT and apply it to obtain the expression for the switching trap model.
\subsection*{Mean first passage time}
We employ a general framework to compute the MFPT for a broad class of random search processes incorporating SR in the presence of single or multiple targets, as developed by Biswas \textit{et al} in Refs. \cite{biswas2024search,biswas2025drift}. The searcher initiates its motion from the origin $O$ in a $d$-dimensional domain and searches for the target; the random time $T$ required to reach the target for the first time is defined as the FPT. However, if the target is not found within a random time $R$ referred as the resetting time during the search phase, the particle is reset to the resetting location stochastically, starting from initial point $x$, referred by the random variable $t_{ret}^O(x)$. Due to the stochastic nature of the return dynamics, the particle may encounter a target during this phase, in which the process terminates at a first-passage event designated by the random variable $t_{ret}^L(x)$, Which means the particle is not blind to the target during the return phase. If the target is not encountered, the particle is reset to the designated resetting position, and the searcher renews it's search dynamics. This renewal cycle continues until the target is found. Taking into account all the above-discussed possibilities, one can formulate a renewal equation by treating $T_R$ the random variable corresponding to the total completion time of the process,
\begin{align}
\begin{array}{l}
T_R = \left\{ \begin{array}{lll}
T, \hspace{3cm} &\text{if }T < R,\\
R + t_{ret}^L(x), \hspace{2cm} &\text{if }R\leq T \hspace{0.2cm}\text{and} ~~t_{ret}^L(x)<t_{ret}^O(x),\\
 R + t_{ret}^O(x)+T_R',\hspace{0.6cm} &\text{if }R\leq T \hspace{0.2cm}\text{and} ~~t_{ret}^O(x)\leq t_{ret}^L(x).   \end{array}\right.\text{ }\end{array}
\label{globalT} 
\end{align}
where $T_R'$ is the identical independent distributed copy of $T_R$. Skipping the intermediate steps of the calculation presented in \cite{biswas2024search,biswas2025drift}, we present the final result below as
\begin{align}
    &\langle T_R \rangle = \frac{\langle \text{min}(T,R) \rangle +\int_\mathcal{D}dx\widetilde{P}_R(x) \langle \text{min}[t_{ret}^L(x), t_{ret}^O(x)] \rangle}{1-\int_\mathcal{D}dx~\widetilde{P}_R(x) \text{Pr}[t_{ret}^O(x)\leq t_{ret}^L(x)]},\label{mfpt_general}
\end{align}
Where $\mathcal{D}$ is the search domain and $\tilde{P}_R(x) = \int_0^{\infty}dt f_R(t)P(x,t)$ is the time integrated propagator of the underlying dynamics $P(x,t)$ in the absence of resetting with waiting time distribution $f_R(t)$. Here,
\begin{align*}
    \langle \text{min}[t_{ret}^L(x), t_{ret}^O(x)] \rangle = ~&\text{Pr}[t_{ret}^L(x)\leq t_{ret}^O(x)]\langle t_{ret}^L(x)|t_{ret}^L(x) \leq t_{ret}^O(x) \rangle  \\ & +\text{Pr}[t_{ret}^O(x)\leq t_{ret}^L(x)]\langle t_{ret}^O(x)|t_{ret}^O(x) \leq t_{ret}^L(x) \rangle
\end{align*}
where $\text{Pr}[t_{ret}^L(x)\leq t_{ret}^O(x)]$ and $\text{Pr}[t_{ret}^O(x)\leq t_{ret}^L(x)]$ denote the splitting probabilities for the particle to reach the target $L$ or the boundary $O$, respectively, starting from position $x$. Similarly, $\langle t_{ret}^L(x)|t_{ret}^L \leq t_{ret}^O(x) \rangle$ and $\langle t_{ret}^O(x)|t_{ret}^O \leq t_{ret}^L(x) \rangle$ denote the conditional MFPTs to reach the target located at $L$ or the boundary $O$, respectively, given that it is reached before the other. For notational simplicity, we henceforth denote $\langle \text{min}[t_{ret}^L(x), t_{ret}^O(x)] \rangle $, $\text{Pr}[t_{ret}^L\leq t_{ret}^O(x)]$ and $\text{Pr}[t_{ret}^O\leq t_{ret}^L(x)]$ as $\langle t^{ret}(x) \rangle$, $\epsilon_O^{ret}(x)$ and $\epsilon_L^{ret}(x)$ respectively. We get,
\begin{align}
    &\langle T_R \rangle = \frac{\langle \text{min}(T,R) \rangle +\int_\mathcal{D}dx\widetilde{P}_R(x) \langle t^{ret}(x) \rangle}{1-\int_\mathcal{D}dx~\widetilde{P}_R(x) \epsilon^{ret}_O(x)},
    \label{mfpt-1d}
\end{align}
For poissonian resetting with an exponential waiting-time distribution $f_R(t) = re^{-rt}$, the expression for $\widetilde{P}_R(x)$,  simplifies to $\widetilde{P}_R(x) = \int_0^{\infty} dt ~re^{-rt} P(x,t) = r\widetilde{P}(x,r)$, where $\widetilde{P}(x,r)$ denotes the Laplace-space propagator of the underlying process. For exponential resetting distribution the expression of $\langle \text{min}(T,R) \rangle$ is given by (discussed in the Refs. \cite{biswas2024search,biswas2025drift})
\begin{align}
    \langle \text{min}(T,R) \rangle = \frac{1 - \widetilde{T}(r)}{r}
\end{align}
where $\widetilde{T}(r)$ is the Laplace transform of first passage distribution function $f_T(t)$. Using the expression of $\langle \text{min}(T,R) \rangle$ and $\widetilde{P}_R(x)$ introduced above, together with the simplified notation adopted earlier, we obtain the following expression for the MFPT,
\begin{align}
    \langle T_R \rangle = \frac{\frac{1 - \widetilde{T}(r)}{r} + r\int_\mathcal{D}dx\widetilde{P}(x,r) \langle t^{ret}(x) \rangle}{[1- r\int_\mathcal{D}dx~\widetilde{P}(x,r) \epsilon^{ret}_O(x)]},
    \label{mfpt_expR}
\end{align}
We now explicitly derive the expression for the MFPT for the switching potential dynamics in the SR case in one dimension. For simplicity, we assume that both the initial and resetting positions are at the origin.\newline
\underline{\textit{MFPT for our model system in one dimension:}}  We now proceed to evaluate all the distinct components entering the general expression for the MFPT in Eq.~\ref{mfpt_expR} above.  The quantity $\langle t^{ret}(x) \rangle$ in Eq. \ref{mfpt_expR} is given by
\begin{align}
    \langle t^{ret}(x) \rangle = \theta(-x)\langle t_1(x) \rangle + \theta(x)\langle t_2(x) \rangle,\label{tret}
\end{align}
It denotes the mean return time under the return dynamics: for $x>0$, it corresponds to the mean time $\langle t_2(x) \rangle$ to reach either the origin $O$ or the target $L$, whereas for $x<0$, it represents the mean time $\langle t_1(x) \rangle$ to reach the origin $O$ from the initial position $x$.

The explicit form of these components appearing in \ref{mfpt_expR} namely, the Laplace transform of the FPT $\tilde{T}(r)$ from Eq.~\ref{LT_fpt}, the Laplace transform of the propagator $\tilde{P}(x,r)$ from Eq.~\ref{LT_prop}, the mean return phase duration when particle starts in $x>0$ domain $\langle t_2(x) \rangle$ from Eq.~\ref{mfpt2}, the mean return time when it starts form $x<0$ domain $\langle t_1(x) \rangle$ from Eq.~\ref{mfpt1}, and the splitting probability to origin $\epsilon_0(x)$ from Eq.~\ref{splitting_origin} have been derived in detail in the Appendices. Substituting these quantities into the MFPT expression $\langle T_r \rangle$ (Eq.~\ref{mfpt_expR}) yields the complete theoretical MFPT for the stochastic resetting with SR in the switching trap model, provided in Appendix~\ref{appA}.
\begin{figure}
    \centering
    \includegraphics[scale=0.5]{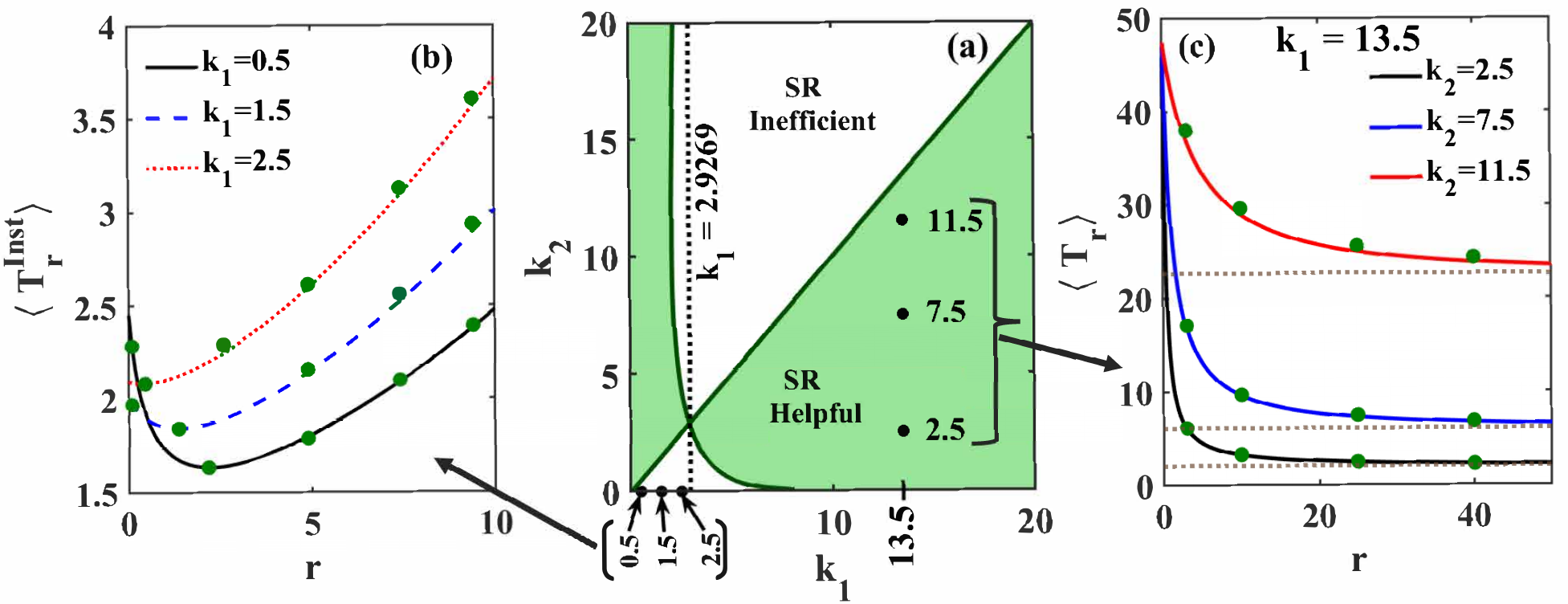}
    \caption{The center figure (a) Phase-space diagram in the $(k_1,k_2)$ plane showing the regions where stochastic resetting with SR is beneficial (green) and the region where it is not beneficial (white), as determined by the CV criterion. The black dotted line separates the region in which instantaneous return is beneficial, corresponding to $k_1<2.9269$. (b) MFPT for IR versus resetting rate $r$ for several values of potential strength in the search phase $k_1$ selected from the regime where IR is beneficial. (c) MFPT versus resetting rate $r$ for stochastic resetting with SR protocol, for different values of $k_2$ at fixed $k_1 = 13.5$, taken from the regime where SR is beneficial. The brown dotted line indicates the limiting value of the MFPT in the large $r$ limit, which corresponds to the MFPT of the underlying OU process with potential strength $k_2$. The green filled circles represent the numerical simulation results in both figures (b) and (c).}
    \label{CVphase}
\end{figure}

\section{Speedup induced by resetting mechanisms}
\label{sec3}
We now identify the region of parameter space in which resetting with SR enhances the efficiency of the search process relative to the underlying dynamics using CV criterion. Furthermore, we also examine whether incorporating SR dynamics leads to improved performance compared to the case of IR.
\subsection{Speed up over underlying process: The CV criterion}
The CV is a statistical concept that originates from probability theory and statistics. It is an indicator of search efficiency in resetting dynamics. It is defined as the ratio of the standard deviation to the MFPT of the reset-free dynamics \cite{dubey2023first,pal2019landau,pal2022inspection,reuveni2016optimal,pal2017first}. 
\begin{align}
    CV=\frac{\sqrt{\langle T^2 \rangle -\langle T \rangle^2}}{\langle T \rangle}=\frac{\sigma(T)}{\langle T \rangle},\label{CVdef}
\end{align}
Here, $\langle T \rangle$ and $\langle T^2 \rangle$ denote the first and second moments of the FPT, respectively, while $\sigma(T)= \sqrt{\langle T^2 \rangle - \langle T \rangle^2}$ representing its standard deviation. In particular, the condition $CV>1$ implies that the resetting with SR enhances the search efficiency relative to underlying dynamics, while $CV<1$ marks regimes where SR is detrimental.\newline
We now investigate the parameter regime in which the resetting with SR process proves to be more effective than the corresponding reset-free dynamics. This implies
\begin{align}
       \langle T_r(\delta r,k_1,k_2) \rangle<\langle T(k_1)\rangle,\label{cv}
\end{align}
where $\langle T\rangle$ is the MFPT of the underlying reset-free process and $ \langle T_r(\delta r, k_1, k_2) \rangle$ is the MFPT in the presence of stochastic resetting with SR is given in Eq. \ref{mfpt_expR}. Expanding the left-hand side of the above equation up to $\mathcal{O}(\delta r)$ and requiring the first-order correction term to be negative yields the following relation (For a detailed derivation, see Ref. \cite{biswas2025drift}).
\begin{align}
    CV^2>  \frac{2}{\langle T \rangle} \overline{\langle t^{ret}(x) \rangle}+2\overline{\epsilon^{ret}_O(x)}-1, \label{cv crit}
\end{align}
This serves as a measure of the relative fluctuations in the FPT of the underlying reset-free process. The overbar $\overline{(...)}$ in Equation~\ref{cv crit} denotes an average taken with respect to the distribution $P_0(x)$, such that $\overline{f(x)}=\int_\mathcal{D}dx~P_0(x)f(x)$. Where the function $P_0(x)$ is defined as,
\begin{align}
    P_0(x) = \frac{1}{\langle T \rangle}\int_0^\infty dt P(x,t)=\frac{1}{\langle T \rangle}\tilde{P}(x,0)
\end{align}
The function $P_0(x)$ is normalized by the construction, ensuring that $\int_\mathcal{D}P_0(x)dx=1$. Eq. \ref{cv crit} represents a universal relation that holds for any stochastic process, irrespective of its underlying dynamics.\newline
We require the first two moments of the FPT for the construction of phase diagram. These moments are obtained from the moment-generating function $\tilde{T}(r)$, defined in Eq.~\ref{LT_fpt}, via the following relation:
\begin{align}
    \langle T^m \rangle =(-1)^m \bigg(\frac{\partial^m \tilde{T}(r)}{\partial r^m}\bigg)\bigg|_{r=0}
\end{align}
The first and second moments of the FPT are given by the following expression,
\begin{align}
   \langle T \rangle = (2k_1)^{-1} \left[2~ _1F_1^{(1,0,0)}\left(0,\frac{1}{2},\frac{k_1}{2}\right) + 2~ H_0^{(1,0)}\bigg(\sqrt{\frac{k_1}{2}}\bigg) + \gamma \right]   \label{1mom}
\end{align}
\begin{align}
  \langle T^2 \rangle = ~\begin{aligned}
       &(8k_1^2)^{-1}\Bigg[ 6\gamma^2 +\pi^2 +16~ H_0^{(1,0)}\bigg(\sqrt{\frac{k_1}{2}}\bigg)^2 +8~H_0^{(2,0)}\bigg(\sqrt{\frac{k_1}{2}}\bigg) + 16~ _1F_1^{(1,0,0)}\Big(0,\frac{1}{2},\frac{k_1}{2}\Big)
       \bigg(\gamma + \\
       & ~ _1F_1^{(1,0,0)}\Big(0,\frac{1}{2},\frac{k_1}{2}\Big) + 8 H_0^{(1,0)}\bigg(\sqrt{\frac{k_1}{2}}\bigg) \bigg( 3\gamma + 4 ~ _1F_1^{(1,0,0)}\Big(0,\frac{1}{2},\frac{k_1}{2}\Big)\bigg) \bigg) - 4~ _1F_1^{(2,0,0)}\Big(0,\frac{1}{2},\frac{k_1}{2}\Big) \Bigg]
   \end{aligned} \label{2mom}
\end{align}
where $\gamma$ is Euler's constant, with numerical value $=0.577216$, $H_n^{(m,o)}(x)$ represents a $m^{th}$ order partial derivative of Hermite polynomial function $H_n(x)$ with respect to its first argument $n$, and $_1F_1^{(m,0,0)}(a,b,z)$ represents a $m^{th}$ order partial derivative of Kummer confluent hypergeometric function $_1F_1(a,b,z)$ with respect to its first argument $a$. To compute the phase-space area in which stochastic resetting with SR is beneficial or detrimental, we substitute the expressions from Eqs.~\ref{LT_prop}, \ref{mfpt2}, \ref{mfpt1}, \ref{splitting_origin}, along with Eqs.~\ref{1mom} and \ref{2mom}, into the criterion given by Eq.~\ref{cv crit}.

The phase-space regions obtained using the CV criterion are shown in Fig.\ref{CVphase} (a). It shows a phase-space diagram identifying the regions where SR is beneficial and where it is detrimental. The green-shaded area corresponds to the regime in which SR reduces the MFPT, while the white region indicates parameter values for which SR is not advantageous over underlying dynamics. The black dotted line separates the domain where IR is beneficial (left of the dotted line) from that where IR is detrimental (right of the dotted line). Although the overall phase space exhibits a complex structure, it contains regions where both instantaneous and stochastic resetting with SR are beneficial, regions where neither protocol is effective, and regions where only one of the two is advantageous. Nevertheless, two distinct subregions can be clearly identified. For large values of $k_1$ and low to moderate values of $k_2$, SR alone is advantageous, as indicated by the green-shaded region. In contrast, for sufficiently small values of $k_1$, corresponding to the region to the left of the black dotted line, IR is favourable. We select representative parameter values from both regions and plot the MFPT for the IR and SR protocols as a function of the resetting rate $r$ in order to validate the phase-space diagram shown in Fig.\ref{CVphase}(b) and (c), respectively.

In Fig.~\ref{CVphase}(b), we plot the MFPT for the IR protocol as a function of the $r$ for different values of $k_1$ chosen from the phase space region where IR is advantageous. For lower values of $k_1$, the MFPT exhibits a non-monotonic dependence on the resetting rate $r$, it initially decreases with increasing $r$, attains a minimum at an optimal resetting rate, and then increases for larger $r$. This behavior can be understood as follows. For small $k_1$, the confining potential is weak, allowing the particle to diffuse relatively freely and explore regions far from the target. Introducing resetting eliminates such long, unfavourable trajectories, thereby reducing the MFPT. However, when the resetting rate exceeds the optimal value, the particle is reset too frequently and does not get sufficient time to reach the target, resulting in an increase in the MFPT. In contrast, for relatively larger values of the potential strength $k_1$, the confining potential strongly localizes the particle near its minimum at the origin. Although thermal fluctuations still enable the particle to reach the target, the process becomes significantly slower. In this regime, resetting interrupts these already rare successful excursions, leading to a monotonic increase of the MFPT with $r$.

Similarly, in Fig.~\ref{CVphase}(c), we present the MFPT for the stochastic resetting with SR protocol for a fixed value of $k_1=13.5$ and three different values of $k_2$. The monotonic decrease of MFPT with $r$ can be understood from the interplay between drift and stochastic resetting. In the search phase, characterised by a strong confining potential ($k_1 = 13.5$ and the other parameters $L = D = \gamma = 1$), the particle experiences a strong drift toward the potential minimum at the origin, which tends to hinder its progress toward the target located at $L$. Although thermal fluctuations allow occasional excursions toward the target, increasing the $r$ increases the chance of reaching the target located at $L$ (in the return phase). Also, in the return phase, the potential strength $k_2 \leq k_1$ results in a comparatively weaker drift toward the origin. This facilitates a higher probability of reaching the target in the return phase due to the reduced confinement and unchanged thermal fluctuations. In the limit of very large resetting rates, the MFPT approaches a finite saturation value, indicated by the brown dashed lines in Fig. \ref{CVphase} (c). This asymptotic value coincides with the MFPT for IR, which is physically intuitive: at sufficiently high resetting rates, the search phase is effectively suppressed, and the dynamics reduce to a particle diffusing and searching for the target solely under the influence of the return phase potential $k_2$. Both figures confirm the conclusions drawn from the phase-space analysis regarding the domains of applicability of the two protocols.

\begin{figure}
    \centering
    \includegraphics[scale=0.51]{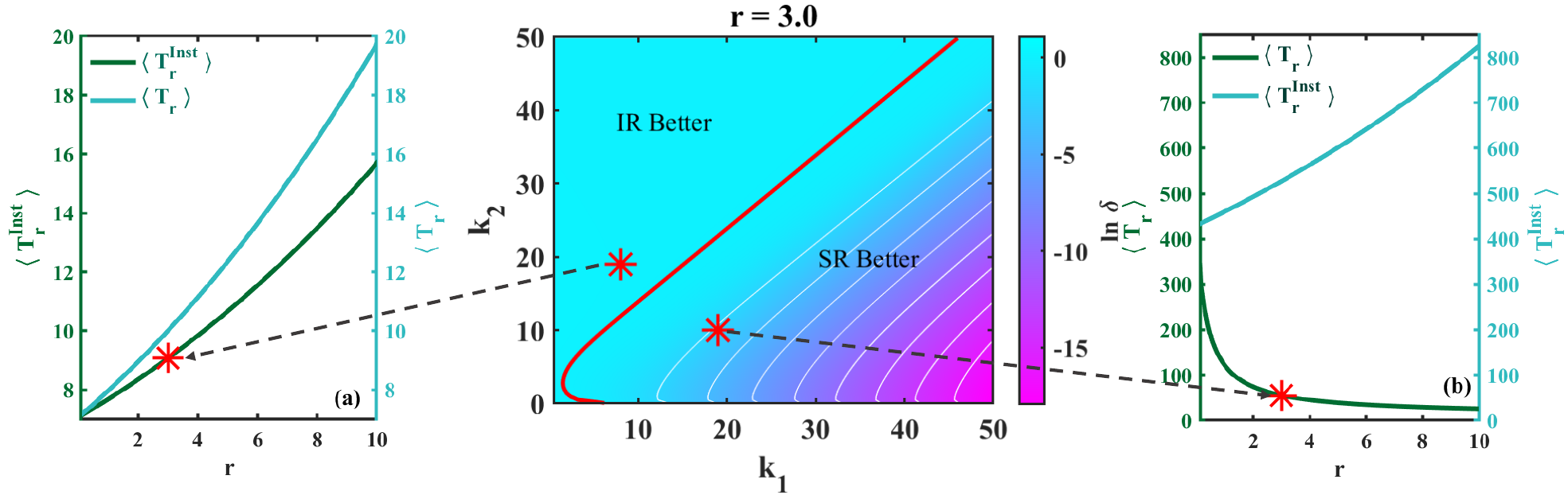}
    \caption{The central figure shows the $(k_1,k_2)$ phase diagram for $r = 3$, highlighting the regime where SR is more efficient than IR. The color scale represents the logarithm of the speed-up parameter $\delta$ defined in \ref{speedup para}. The red curve denotes the separatrix between the two regions corresponding to $\delta = 1$. The outer panels illustrate representative MFPT behavior for selected points in the phase space. In panel (a), the MFPT is shown as a function of the resetting rate $r$ for the parameter $(k_1,k_2) = (8,19)$, chosen from the region where IR is advantageous. The green line corresponds to the MFPT for the IR protocol, while the blue line represents the MFPT for the SR protocol at the same parameter values, demonstrating that IR yields a lower MFPT in this regime. In panel (b), the MFPT is plotted as a function of $r$ for the SR protocol for the parameter set $(k_1,k_2) = (19,10)$, selected from the region where SR is advantageous, illustrating the improved performance of the MFPT with $r$ in this regime. The green line  corresponds to MFPT for SR protocol, and the blue is for the instantaneous return protocol. The results shown in panels (a) and (b) confirm the predictions of the phase diagram, demonstrating that the selected parameter sets exhibit the expected MFPT behavior in the respective IR- and SR-favorable regions.}
    \label{PhaseSR}
\end{figure}

\subsection{Speedup over instantaneous resetting—the stochastic return (SR) criterion} 
We now examine the effect of SR relative to IR using the SR criterion.
SR affects the overall search time in two main ways when compared with IR. First, unlike IR, which brings the searcher back to the origin without any time cost, SR requires a finite time to complete. Second, because the return is stochastic, the searcher may encounter the target during the return phase itself, thereby completing the search without restarting. This possibility does not exist for IR, where the searcher is returned immediately to the origin and the search always begins anew. As a result, SR introduces a trade-off between the extra time spent returning and the increased chance of finding the target during return. To quantify this trade-off, we compare the MFPTs for the two return protocols (i.e., SR and IR protocols) and determine the condition under which SR outperforms IR, i.e., mathematically we can express it as
\begin{align}
    \langle T_r \rangle < \langle T^{\text{Inst}}_r \rangle\label{cond},
\end{align}
Using Eq.~\eqref{cond}, we write a general criterion for identifying the phase-space regime in which SR is advantageous, discussed in Ref. \cite{biswas2025drift}. 
\begin{align}
    \mathcal{T} \equiv \frac{\langle \text{min}[t_{ret}^L, t_{ret}^O] \rangle}{\text{Pr}[t_{ret}^L< t_{ret}^O]}~~ < ~~ \langle T_r^{\text{Inst}} \rangle \label{SR crit}
\end{align}
where $\text{Pr}[t_{ret}^L< t_{ret}^O] = \int_\mathcal{D}dx\widetilde{P}_R(x) \text{Pr}[t_{ret}^L(x)< t_{ret}^O(x)]$ and $\langle \text{min}[t_{ret}^L, t_{ret}^O] \rangle =\int_\mathcal{D}dx\widetilde{P}_R(x) \langle \text{min}[t_{ret}^L, t_{ret}^O] \rangle$, with $\mathcal{D}$ being the search domain. In one dimension, the resulting criterion for exponential resetting takes the following form:
\begin{align}
    \mathcal{T}\equiv \frac{\int_{-\infty}^1dx\widetilde{P}(x,r)\langle t^{ret}(x) \rangle}{\int_{0}^1dx\widetilde{P}(x,r) \epsilon_L(x)}<\langle T^{\text{Inst}}_r \rangle, 
     \label{sr crit}
\end{align}
where $\epsilon_L(x)$ denotes the probability that the searcher reaches the target, starting from $x>0$ during the return phase, and other quantities have been defined previously. The quantity $\mathcal{T}$ captures the trade-off between the return time and the probability of locating the target during the stochastic return. In particular, the numerator of $\mathcal{T}$ in Eq.~\ref{sr crit} accounts for the return time, larger return times increase $\mathcal{T}$ and thereby making the criterion harder to satisfy. In contrast, the denominator represents the probability of finding the target during the return phase; higher values reduce $\mathcal{T}$ and facilitate the fulfillment of the criterion in Eq.~\ref{sr crit}. The left-hand side of Eq.~\ref{sr crit} is evaluated by substituting Eqs.~\ref{LT_prop}, \ref{mfpt2}, \ref{mfpt1}, and \ref{splitting_target} into Eq.~\ref{sr crit}.\newline
In Eq. \ref{cond}$, \langle T^{Inst}_r \rangle$ denotes the MFPT corresponding to IR and it can be obtained from Eq. \ref{mfpt_expR} by noting that the return times $\langle t_1(x) \rangle$ and $\langle t_2(x) \rangle$ vanish, since the particle is returned to the origin instantaneously. Moreover, because the particle reaches the origin with certainty, the splitting probability to the origin is unity, i.e., $\epsilon_O(x) =1$. Combining these observations, we obtain (for a detailed derivation see Ref. \cite{biswas2025drift})
\begin{align}
    \langle T^{\text{Inst}}_r \rangle=\frac{1-\widetilde{T}(r)}{r\widetilde{T}(r)},\label{MFPT_Inst}
\end{align}
We substitute $\tilde{T}(r)$ from Eq. \ref{LT_fpt} into Eq.~\ref{MFPT_Inst} yields the MFPT for IR:
\begin{align}
    \langle T^{\text{Inst}}_r \rangle = \frac{D}{L^2}\frac{1}{r}\Bigg[\frac{\pi H\left( -\frac{r}{k_1},\sqrt{\frac{k_1}{2}}\right)- \Gamma\left(\frac{r}{2k_1}\right)\Gamma\left(1-\frac{r}{k_1}\right)~_1F_1\left(\frac{r}{2k_1},\frac{1}{2},\frac{k_1}{2}\right)\sin\left(\frac{\pi r}{k_1}\right)}{\frac{2^{-\frac{r}{k_1}}\pi^{\frac{3}{2}}}{\Gamma\left( \frac{1}{2} +\frac{r}{2k_1} \right)} -\Gamma\left(\frac{r}{2k_1}\right)\Gamma\left(1-\frac{r}{k_1}\right)\sin\left(\frac{\pi r}{k_1}\right)}-1\Bigg],\label{mfptIR_1d}
\end{align}
Substituting LHS and RHS into Eq.~\ref{sr crit} yields the phase diagram, which delineates the parameter regimes where stochastic resetting with SR enhances or hinders the search process relative to IR. We also define a speed-up parameter to quantify the advantage of stochastic resetting with SR over IR.
\begin{align}
    \delta=\frac{\langle T^{\text{Inst}}_r \rangle}{ \langle T_r \rangle} \label{speedup para}
\end{align}
The stochastic resetting with SR strategy outperforms IR for $\delta>1$, with the value of $\delta$ serving as a measure of the speed-up. The boundary separating the two regimes is obtained by setting $\delta=1$ in \ref{speedup para}. The expression for $\delta$ can be evaluated exactly using Eq. \ref{mfptIR_1d} and Eq. \ref{mfptSR_1d} in Eq.\ref{speedup para}. We now construct the phase-space plots and identify the regions where IR or SR is advantageous, based on the criteria discussed above.

\textbf{\underline{Description of phase space plot:}} The parameter spaces spanned by $(k_1,k_2)$ for a fixed value of $r=3.0$ is shown in central Fig.~\ref{PhaseSR}. The phase diagrams are constructed using the SR criterion defined in Eq. \ref{sr crit}. The color gradient represents the magnitude of the speed-up parameter $\delta$. The separatrix, indicated by the bold red line, marks the boundary between the regimes where IR and SR  dominate, i.e., on this line the two protocols are equivalent ($\delta=1$). To the left of the red line, IR is favourable, while to the right stochastic resetting with SR yields a lower MFPT.

We now examine the behavior of the phase-space region. Mostly for a fixed resetting rate $r$ and a given value of $k_1$, SR is advantageous when $k_2 \lesssim k_1$. In this regime, the return potential is relatively weak, allowing the searcher to diffuse more freely and to detect the target during the return phase, which leads to a reduction in the MFPT. As $k_2$ increases, the searcher experiences a stronger restoring force toward the resetting location (the origin in the present case), which suppresses its ability to reach the target.

To illustrate the two regimes, we select one representative point (marked by a red asterisk) from each region of the phase diagram: $(k_1,k_2)=(19,10)$ from the stochastic return favorable regime and $(k_1,k_2)=(8,19)$ from the IR favorable regime. Figure~\ref{PhaseSR}(a) shows the MFPT as a function of the resetting rate for the representative point in the IR favorable regime. The comparison demonstrates that IR consistently yields a lower MFPT than SR over the entire range of resetting rates considered. Conversely, Fig.~\ref{PhaseSR}(b) presents the MFPT for the representative point in the SR favorable regime, where SR consistently outperforms IR for all resetting rates.
\section{Experimental protocols for stochastic resetting WITH SR}
\label{sec4}
We now consider several experimental realizations of the switching-potential protocol. We use criterion \ref{SR crit} to construct the phase-space diagram, thereby identifying the region in which SR outperforms IR. The system is characterized by three control parameters, the resetting rate and the potential strengths in the search and return phases. The phase diagram is generated by varying two of these parameters while keeping the third fixed. 

We now consider two types of experiments: one in which a switching potential model, as used in our framework, is employed, and another in which the search phase is purely diffusive while the return phase occurs in the presence of a confining potential. Using experimentally relevant parameter values, we project the corresponding points onto the phase diagram to determine whether they lie in the SR favourable or IR favorable regime.
\subsection{Switching potential between Search and Return Dynamics}
\begin{figure}
    \centering
    \includegraphics[scale=0.4]{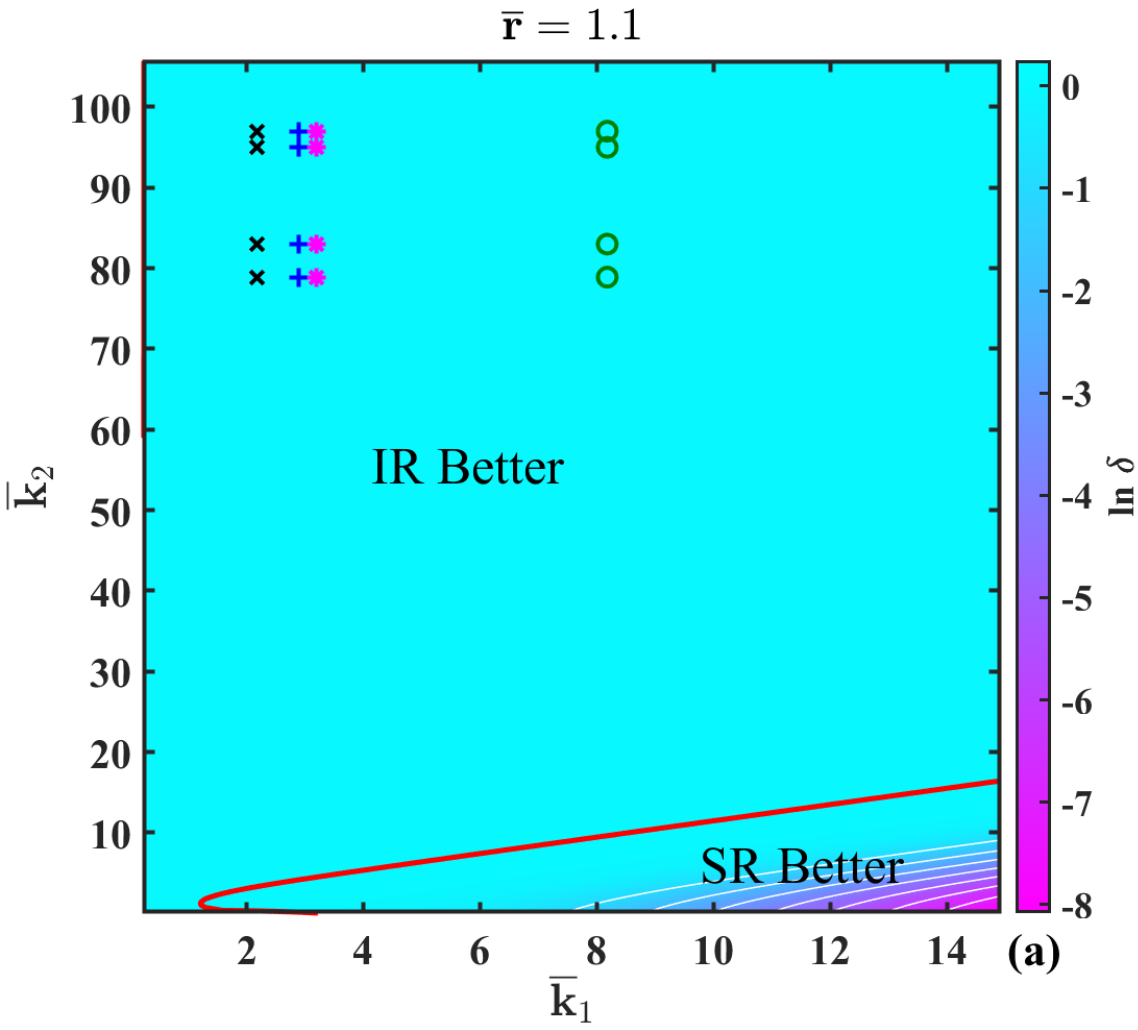}\hspace{0.3 cm}
    \includegraphics[scale=0.4]{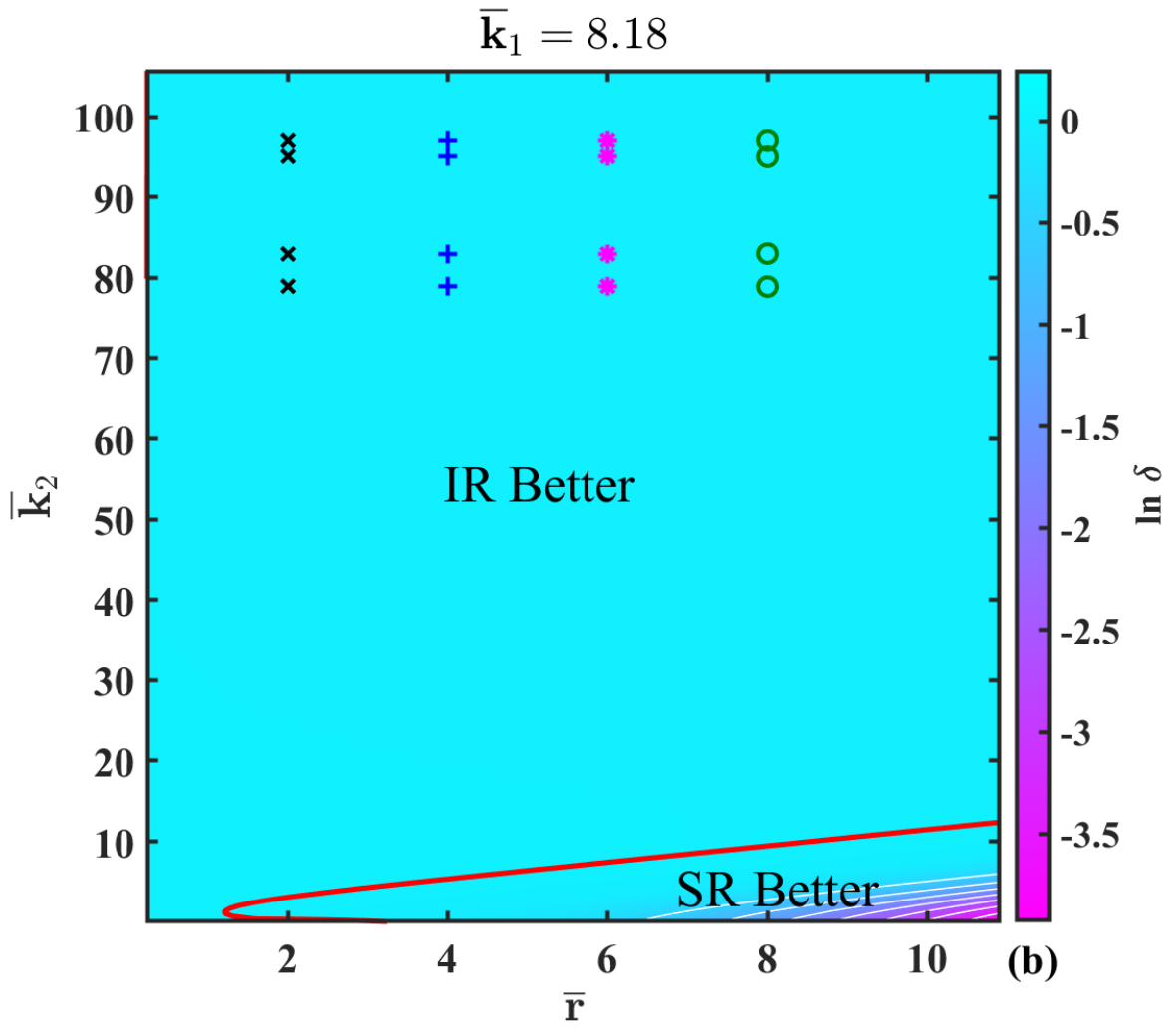}
    \caption{(a) $(\overline{k}_1,\overline{k}_2)$ phase space for fixed $\overline{r}=1.1$ and (b) $(\overline{r},\overline{k}_2)$ phase space for fixed $\overline{k}_1=8.18$. The color gradient represents the logarithm of the speed-up parameter $\delta$. The red line ($\delta=1$) marks the separatrix between the regimes where IR and SR are more efficient. The region with $\delta<1$ corresponds to parameters for which SR is advantageous (i.e., the SR criterion is satisfied), whereas the region with $\delta>1$ corresponds to parameters where IR performs better. In both panels, the symbols denote the parameter values adopted from Ref. \cite{goerlich2025taming}.}
    \label{RemiExp}
\end{figure}
Here, we discuss two experiments that implement the newly designed switching potential  SR protocol, one by Goerlich \textit{et al.} \cite{goerlich2025taming} and another by Biroli \textit{et al.} \cite{jkn8-h939}. In Ref. \cite{goerlich2025taming}, the paper aims to experimentally realize stochastic return in a controlled microscopic system to demonstrate its Maxwell's-demon like behavior and to quantify its associated energetic and informational costs. The potential strength in the search phase ranges from 2.18 to 8.18 ~$\mathrm{pN}/\mu\mathrm{m}$, while in the return phase it varies between 78.9 and 97 $~\mathrm{pN}/\mu\mathrm{m}$. The resetting rate used in the experiment lies in the range 164 to 790 $\mathrm{s}^{-1}$.

Here, extremely high resetting rates were employed to probe the steady-state properties of the system. However, in the context of first-passage phenomena, such large values of $r$ are not appropriate. At very high resetting rates, the particle is reset so frequently that it does not have sufficient time to reach the target before being reset, thereby suppressing the first-passage process. Therefore, in our analysis, we restrict ourselves to moderate and physically relevant values of $r$, consistent with those accessible in experiments, in order to properly capture the first-passage dynamics. In the phase space plot, we have used values of the resetting rate $r$ in the range $0$ to $11~\mathrm{s}^{-1}$.

The other relevant parameters used to construct the phase space plot, like the temperature of the surrounding medium is $T = 296~\mathrm{K}$, the friction coefficient $\gamma = 0.026~\mathrm{pN\,s}/\mu\mathrm{m}$ and hence the diffusion constant $D = k_B T/\gamma = 0.16~\mu\mathrm{m}^2/\mathrm{s}$ . Since the experimental parameters are reported in dimensional units, we also present the phase diagram in dimensional variables to enable a direct comparison with the experiments, as shown in Fig.~\ref{RemiExp}.

\begin{figure}
    \centering
    \includegraphics[scale=0.4]{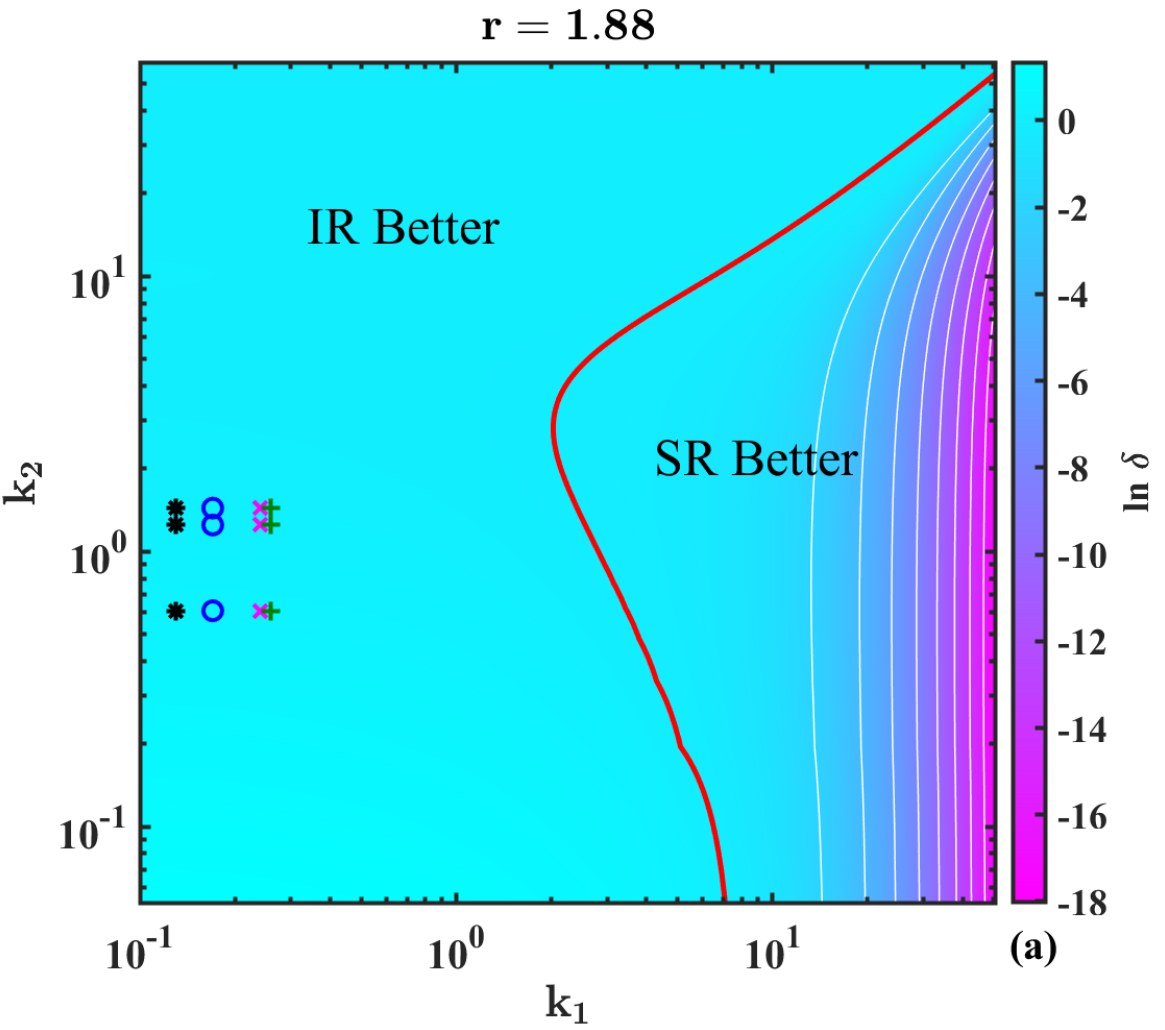}\hspace{0.3cm}
    \includegraphics[scale=0.4]{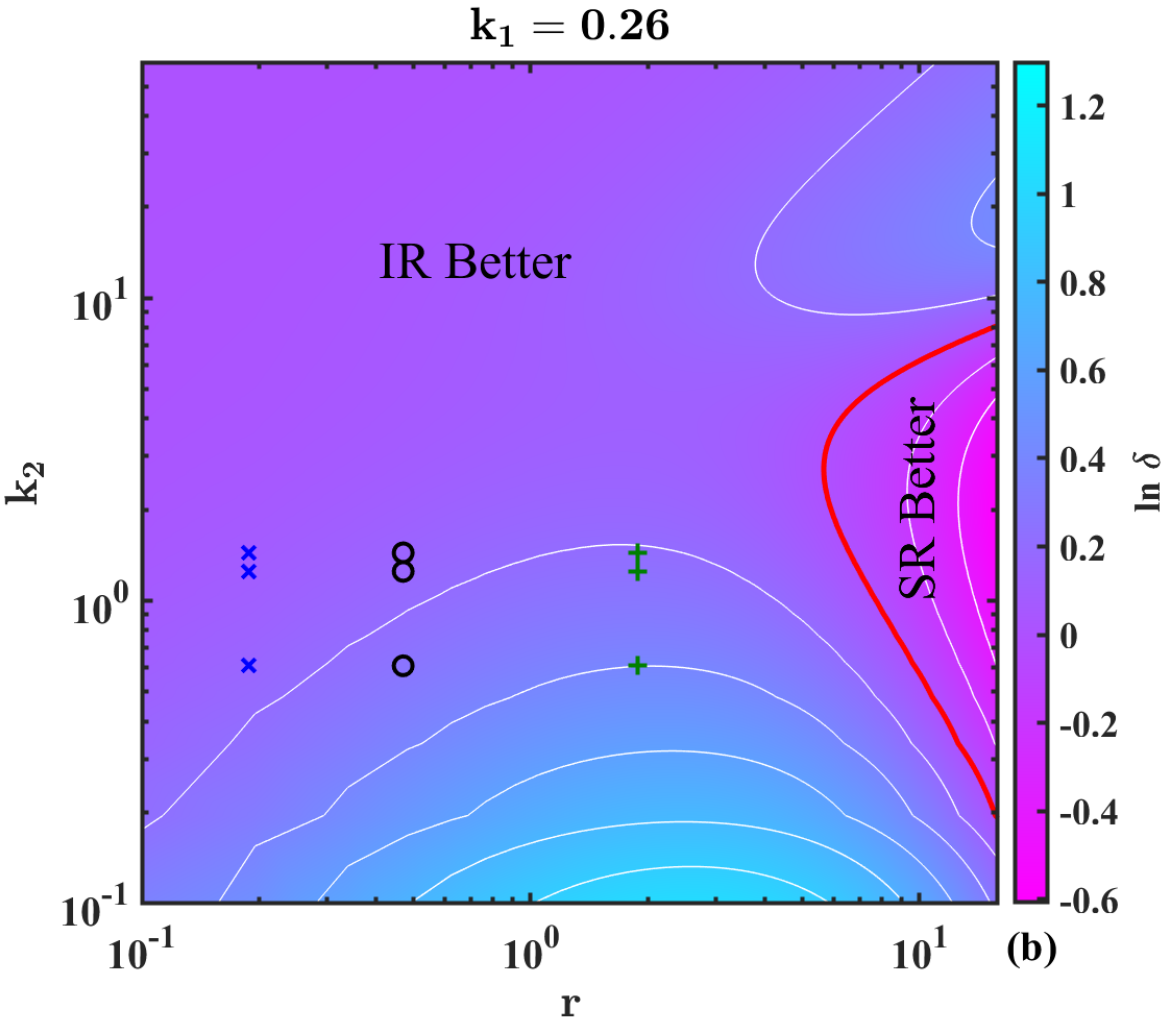}
    \caption{(a) $(k_1,k_2)$ phase space for fixed $r=1.88$ and (b) $(r,k_2)$ phase space for fixed $k_1=0.26$. The color gradient represents the logarithm of the speed-up parameter $\delta$. The red line ($\delta=1$) indicates the separatrix separating the IR better and SR better regions. The SR better region corresponds to parameter values for which the SR criterion is satisfied, whereas the IR better region corresponds to those for which it is not. In both figures, the symbols denote the parameter values adopted from Ref. \cite{jkn8-h939}.}
    \label{BiroliExp}
\end{figure}

In Ref. \cite{jkn8-h939}, recently Biroli \textit{et al.} also investigated a closely related model consisting of $N$ independent Brownian particles evolving in a switching harmonic potential. In this study, the potential strength ranges from 0.13 to 0.26 in the search phase and from 0.61 to 1.44 in the return phase, and $r$ ranges from 0.188 to 1.88 are used. Since the system obeys the rescaled form of the Langevin equation, the phase-space plots are presented entirely in terms of these dimensionless variables shown in Fig. \ref{BiroliExp}.
\newline
\textbf{\textit{Description of phase space diagram: }} We note that the authors in Refs. \cite{jkn8-h939,goerlich2025taming} did not investigate first-passage properties. In contrast, we utilize the experimental parameters reported in their study to analyze the MFPT within our theoretical framework. In particular, we map these parameters onto the phase space of our model to identify the regimes where IR or SR is more effective. In Fig. \ref{RemiExp} and \ref{BiroliExp}, we have plotted the $(k_1,k_2)$ phase space region for a fixed value of $r$ and $(r,k_2)$ the phase space region for a fixed value of $k_1$. In both figures, the different symbols denote the experimental parameter values used in Refs. \cite{goerlich2025taming} and \cite{jkn8-h939}, respectively. It is evident that the experimental parameter values lie within the region of parameter space where only IR is advantageous. However, extrapolation of the phase space in both figures reveals a regime in which stochastic resetting also becomes beneficial. The red line denotes the separatrix obtained from the condition $\delta=1$ in Eq. \ref{speedup para}, which separates the region where the SR criterion is satisfied (implying that SR is beneficial i.e. $\delta<1$) from the region where the SR criterion is not satisfied (indicating that IR is favorable i.e. $\delta>1$). The color gradient represents the logarithm of the speed-up parameter $\delta$.

Beyond the switching trap-based experiments discussed above, we also consider another widely discussed class of experimental realizations wherein the search phase is governed by free diffusion rather than a trapping potential, while the return phase is implemented using a harmonic trap. 
\subsection{Free Diffusive Search ($k_1=0$) with Potential-Guided Return}
In Refs.~\cite{besga2020optimal,besga2021dynamical}, Besga \textit{et al.} present both experimental and theoretical investigations of the optimal MFPT for a Brownian searcher reaching a target under SR. They consider two types of resetting protocols: periodic and random (Poissonian) resetting. The key difference between their model and ours lies in the search phase and the nature of resetting. In their case, the search phase corresponds to pure diffusion ($k_1 = 0$), and the resetting is instantaneous. In contrast, our model incorporates a switching-potential protocol with a finite return phase. To establish a connection with their experimental setup, we take the parameter values reported in their work and map them onto our SR framework by setting $k_1 = 0$. We then analyze how these parameters are positioned within our phase-space diagram.

The experimental parameters are mapped onto our theoretical model as follows discussed in detail in Ref.~\cite{biswas2025drift}. The potential strengths $k_2$ in the theoretical description are related to the experimentally realized trap stiffness via a relation
\begin{align}
    k_2 = b^2
\end{align}
Where $b=L/\sigma$, here, $L$ denotes the distance of the target from the trap center, and $\sigma=\sqrt{\frac{k_B T_B}{\kappa}}$, represents the standard deviation of the equilibrium position distribution of the particle. In these experimental studies, the parameter $b$ is varied in the range 2 to 3, which corresponds in our model to a return-phase potential strength $k_2$ ranging from 4 to 9.\newline
 The resetting rate in the theoretical model is related to its experimental counterpart through the following relation.
 \begin{align}
    r = c^2
\end{align}
Where $c$ is the dimensionless resetting parameter defined by $c = \sqrt{\frac{r}{D}}L$. The dimensionless experimental parameter $c$ varies in the range 0 to 14, which in our model corresponds to $r$ ranging from 0 to 196.
\begin{figure}
    \centering
    \includegraphics[scale=0.8]{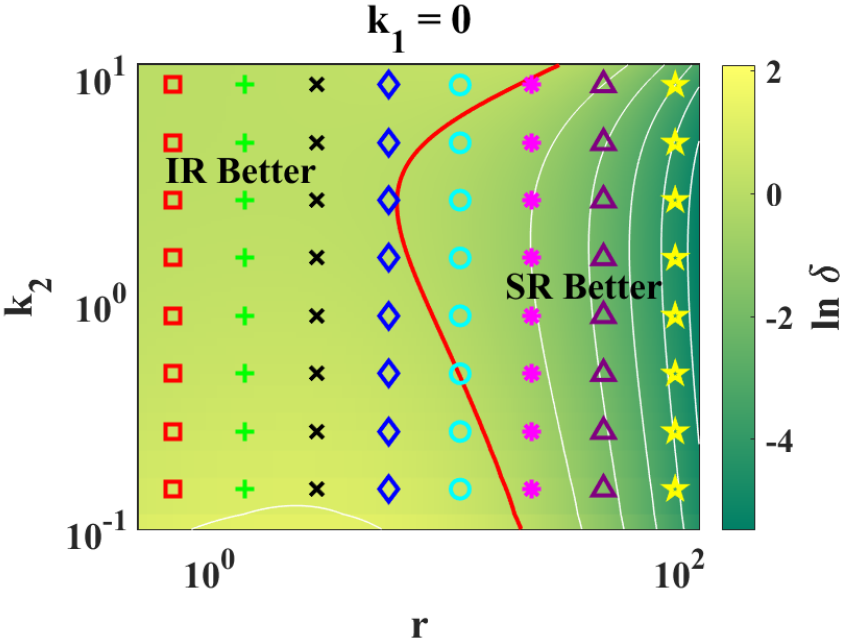}
    \caption{$(r,k_2)$ phase space for $k_1=0$ corresponding to a purely diffusive search process with SR implemented via a harmonic trapping potential. The red line ($\delta=1$) indicates the separatrix separating the IR-better and SR-better regions. The SR-better region corresponds to parameter values for which the SR criterion is satisfied, whereas the IR-better region corresponds to those for which it is not. The symbols mark selected parameter points, including values adopted from the experimental study in Ref. \cite{besga2020optimal,besga2021dynamical}.}
    \label{Exp3}
\end{figure}
\newline
Figure~\ref{Exp3} shows the $(r,k_2)$ phase diagram for the case of purely diffusive search ($k_1 = 0$). We consider the range of values of $r$ and $k_2$ used in the experimental studies and map them onto the phase-space diagram. We find that some of these parameter sets lie in the regime where SR is more efficient, while others fall within the regime where IR performs better.

We now summarize the experimental aspects of our study and connect them with the theoretical framework developed in the previous sections. We demonstrate that stochastic resetting with SR provides a more realistic experimental framework than the idealized IR protocol for investigating the first-passage properties of physically realizable systems. In contrast to the IR protocol, SR does not require the relaxation segments to be removed from the recorded trajectories, thereby allowing the first-passage dynamics to be studied directly from the complete experimental data. In the initial part of the theoretical analysis, we explored a broad parameter space to identify the regimes in which SR outperforms IR in terms of first passage performance for switching trap model (Fig. \ref{PhaseSR}). In the experimental section, we consider parameter sets from existing experiments on similar switching-trap models and map them onto our theoretical phase diagram. We find that the available set of experimentally relevant parameters lie predominantly in the region where IR outperforms SR (Fig. \ref{RemiExp}, \ref{BiroliExp}). We also considered parameter values from experiments involving diffusive search followed by OU return and mapped them onto our theoretical phase diagram. Interestingly, these parameter sets span both regimes, with some falling in the region where SR outperforms IR, while others lie in the regime where IR outperforms SR (Fig. \ref{Exp3}). In these experiments \cite{goerlich2025taming,jkn8-h939,besga2020optimal,besga2021dynamical}, a high trap stiffness during the return phase was chosen to approximate IR. In contrast, our model imposes no such restrictions on the key parameters, including the trap stiffness during the search and return phases and the resetting rate, thereby allowing systematic exploration of the full parameter space.
Our results therefore suggest that a broad range of parameter values can now be explored experimentally, enabling access to different regions of the phase space and facilitating a systematic investigation of the regimes in which SR or IR yields superior first-passage performance.

\section{Conclusions}\protect \label{sec5}
In this work, we studied a stochastic search process with stochastic resetting with SR, which means the return to the resetting position occurs through a finite-time stochastic return rather than an instantaneous jump. Both the search and return phases were modeled using the OU process, allowing us to capture a realistic scenario where the particle may still detect the target during the return phase. We derive an analytical expression for the MFPT for the switching trap model in one dimension and identify parameter regimes where stochastic resetting with SR enhances or hinders search efficiency compared to underlying dynamics and construct the corresponding phase diagram using the CV criterion. We also compare the SR with the IR using the SR criterion and identified the parameter regimes in which SR outperforms IR. We find that SR is most effective in optimizing the search in the region characterized by large $k_1$ and low to moderate values of $k_2 \le k_1$. 

The main goal of this work is to establish a direct connection between our theoretical framework and existing experimental realizations of resetting. To this end, we consider switching-trap dynamics in which distinct potential traps govern the search and return phases. This framework shares several qualitative features with existing experiments, particularly in the implementation of the return phase. But the search dynamics vary across implementations; some employ free diffusion, whereas others involve diffusion in a confining potential, as considered in our model. At the same time, our model allows for a finite and stochastic return duration, providing a natural extension beyond the idealized IR picture. In many experimental studies, IR is approximated by constructing trajectories in which the relaxation segments following a reset are removed from the recorded dynamics. Our framework, in contrast, retains the return dynamics explicitly and allows the stiffness of the return potential to take arbitrary values, without requiring the return trap to be infinitely stiff. An important consequence is that the target remains accessible throughout the return phase, so that a first passage can occur either during the search phase or while the particle is returning toward the reset position. We also consider resetting to a fixed position, whereas experimentally realized protocols may involve reset positions sampled from the equilibrium distribution. These differences notwithstanding, the switching-trap framework provides a common setting in which the two protocols can be compared and related to experimentally accessible dynamics.

To establish a connection with existing experiments, we map parameter values reported in the literature onto our theoretical phase-space diagram. The available experimental parameter sets provide useful reference points and, as shown in the phase diagram, predominantly fall within the regime where IR is more efficient. Importantly, however, our theoretical analysis reveals a broader parameter space in which SR can outperform IR. We therefore view the phase-space diagram not as a prescription for existing experimental protocols, but as a guide toward unexplored regimes in which the consequences of finite-time stochastic return may become experimentally observable. Exploring these regimes by varying experimentally accessible quantities such as the trap stiffnesses, switching rates, and resetting rates could provide a direct test of our theoretical predictions and help assess the practical advantages of SR. We hope that this connection between theory and existing experimental parameter ranges will motivate further experimental investigations of switching-trap resetting beyond the IR limit. More importantly, our framework enables realistic experiments in which the measured MFPT can be directly compared with theoretical predictions for stochastic return, without removing the return segments from recorded trajectories.

The study of resetting with stochastic return is still at an early stage, leaving several directions for further exploration. A natural extension is to combine SR with non-Poissonian resetting, a protocol also investigated experimentally \cite{besga2020optimal}. Another promising direction is to examine the energetic costs associated with the switching-trap model within the framework of stochastic thermodynamics \cite{tal2020experimental,gupta2020work}. Quantifying the work involved in switching the trap potentials and the dissipation during the return phase could reveal trade-offs between search efficiency and energetic expenditure, and help identify protocols that balance speed and cost.

It would also be interesting to combine stochastic return with threshold resetting \cite{de2020optimization,752c-wqly,biswas2026optimal,das2026non}, where the initiation of resetting is triggered by the particle reaching a prescribed threshold rather than by a random resetting time. Such extensions, together with a thermodynamic characterization of the resetting and return phases, could provide a deeper understanding of the interplay between first-passage performance and energetic cost in physically realizable resetting protocols.

\begin{acknowledgments}
AD acknowledges support from the IMSc Post-Doctoral Fellowship. AD acknowledges Arup Biswas for fruitful discussions during the course of this work. We thank Rahul Das for assisting us in designing the schematic figure. AP acknowledges research funding under the scheme ANRF/ARGM/2025/001623 from ANRF, India and research support from the Department of Science and Technology, India, SERB Start-up Research Grant Number SRG/2022/000080. AP also acknowledges the International Research Project (IRP) titled “Classical and quantum dynamics in out of equilibrium systems” by CNRS, France. Finally, we gratefully acknowledge research support from the Department of
Atomic Energy, Government of India via Soft Matter Apex projects.
\end{acknowledgments}

\appendix

\section{Mean First Passage Time Results for the Switching Trap Model}\label{appA}
Here, we present the expression for the MFPT $\langle T_r \rangle$ of the switching-trap model under the SR protocol, which is given by
\begin{align}
    \langle T_r \rangle = \frac{L^2}{D}\small\frac{\begin{aligned}
    &\left(1-  \frac{\frac{2^{-\frac{r}{k_1}}\pi^{\frac{3}{2}}}{\Gamma\left( \frac{1}{2} +\frac{r}{2k_1} \right)} -\Gamma\left(\frac{r}{2k_1}\right)\Gamma\left(1-\frac{r}{k_1}\right)\sin\left(\frac{\pi r}{k_1}\right) }{\pi H\left( -\frac{r}{k_1},\sqrt{\frac{k_1}{2}}\right)- \Gamma\left(\frac{r}{2k_1}\right)\Gamma\left(1-\frac{r}{k_1}\right)~_1F_1\left(\frac{r}{2k_1},\frac{1}{2},\frac{k_1}{2}\right)\sin\left(\frac{\pi r}{k_1}\right)}\right )\\
    &+r^2\int_{-\infty}^1 dx~\pmb{\Bigg[}\frac{
       e^{-\frac{k_1x^2}{2}}\left[-x\theta(x)~_1F_1\left(\frac{k_1+r}{2k_1},\frac{3}{2},\frac{k_1x^2}{2}\right) +~ _1F_1\left(\frac{k_1+r}{2k_1},\frac{3}{2},\frac{k_1}{2}\right)\pmb{N_1}\right]}{\Gamma\left(\frac{r}{2k_1}\right)~ _1F_1\left(\frac{r}{2k_1},\frac{1}{2},\frac{k_1}{2}\right) +\sqrt{2k_1}\Gamma\left(\frac{k_1 + r}{2k_1}\right)~_1F_1\left(\frac{k_1+r}{2k_1},\frac{3}{2},\frac{k_1}{2}\right)} \\ 
       &\times\Bigg[ \theta(-x) \frac{1}{2}\left( \frac{\pi}{k_2}\text{Erfi}\Big(\sqrt{\frac{k_2}{2}}x\Big)\left( 1 - \text{Erfi}\Big(\sqrt{\frac{k_2}{2}}x\Big) \right) +  x^2~ _2F_2\left((1,1),\Big(\frac{3}{2},2\Big),-\frac{k_2x^2}{2}\right)\right) +\theta(x)\\
        &\times \frac{1}{2} \left( \frac{\pi}{k_2} \text{Erfi}\left(\sqrt{\frac{k_2}{2}}x \right) \left( \text{Erf}\Big(\sqrt{\frac{k_2}{2}}\Big)-\text{Erf}\Big(\sqrt{\frac{k_2}{2}}x\Big) \right) + x^2~ _2F_2\left((1,1),\Big
        (\frac{3}{2},2\Big),-\frac{k_2x^2}{2}\right) - ~\pmb{N_2} \right)     \Bigg] \pmb{\Bigg]} 
    \end{aligned}}{\tiny r\left[1-r\displaystyle\int_{-\infty}^1dx\left( \frac{e^{-\frac{k_1x^2}{2}}\left(-x\theta(x)_1F_1\left(\frac{k_1+r}{2k_1},\frac{3}{2},\frac{k_1x^2}{2}\right) +~ _1F_1\left(\frac{k_1+r}{2k_1},\frac{3}{2},\frac{k_1}{2}\right)\pmb{N_1}\right)}    
    {\Gamma\left(\frac{r}{2k_1}\right)~ _1F_1\left(\frac{r}{2k_1},\frac{1}{2},\frac{k_1}{2}\right) +\sqrt{2k_1}~\Gamma\left(\frac{k_1 + r}{2k_1}\right)~_1F_1\left(\frac{k_1+r}{2k_1},\frac{3}{2},\frac{k_1}{2}\right)}\left( 1-\frac{\text{Erfi}\Big(\sqrt{\frac{k_2}{2}}x\Big)}{\text{Erfi}\Big(\sqrt{\frac{k_2}{2}}\Big)}\right)\right ) \right]},\label{mfptSR_1d}
\end{align}
Where $N_1 = \left[ \Gamma\left(\frac{r}{2k_1}\right)~ _1F_1\left(\frac{r}{2k_1},\frac{1}{2},\frac{k_1x^2}{2}\right) +\sqrt{2k_1}~x \Gamma\left(\frac{k_1+r}{2k_1}\right)~ _1F_1\left(\frac{k_1+r}{2k_1},\frac{3}{2},\frac{k_1x^2}{2}\right)\right]$,\newline
$N_2 = \frac{\text{Erfi}\left(\sqrt{\frac{k_2}{2}}x\right)~_2F_2\left((1,1),(\frac{3}{2},2),-\frac{k_2}{2}\right)}{\text{Erfi}\left(\sqrt{\frac{k_2}{2}}\right)}$. where $\theta(x)$ is the Heaviside step function defined as $\theta(x)=0$ for $x<0$ and $\theta(x)=1$ for $x>0$, $H(n,x)$ is Hermite polynomials, $\text{Erfi}(z)=\text{erf}(iZ)/i$ is the imaginary error function and $i=\sqrt{-1}$ and $\text{erf}(z)$ is the error function, $\Gamma(z)$  is the Euler gamma function, $_1F_1(a,b,z)$ the Kummer confluent hypergeometric function and $_pF_q[(a_1,...,a_p),(b_1,...,b_q),z]$ is the generalized hypergeometric function. The remaining symbols appearing in the above equation have been defined earlier.  This expression gives the MFPT for the switching potential model with $k_1$ and $k_2$ denote the potentials in the search and return phases, respectively, $r$ is the exponentially distributed resetting rate, and the searcher starts from the origin in one dimension. Since the integral cannot be evaluated in closed form, we compute it numerically using Mathematica to obtain the MFPT.

\section{Derivation of Laplace Transform of FPT distribution function $\tilde{T}(r)$}\label{appB}
To compute $\tilde{T}(r)$, we begin by introducing a first-passage Brownian functional, defined over a random time interval $[0,t_f]$ (see Ref.\cite{majumdar2005brownian})
\begin{align}
Z = \int _{0}^{t_f} U[x(\tau)] d \tau~,
\label{FPBF}
\end{align}
where $t_f$ denotes the first-passage time, which is a random variable, and $x(\tau)$ denotes a Brownian trajectory starting from $x_0$ at $\tau=0$ and continuing until it hits the target boundary at the first-passage time $\tau=t_f$. $U(x)$ is some specified function and the form depends on the quantity we are interested in. Since $Z$ is a random quantity, a better quantity is to calculate its distribution function $f_Z(x_0)$. As our focus is on first-passage functionals, we begin by taking the Laplace transform of the 
\begin{align}
\tilde{T}(r,x_0) &=\int _{0}^{\infty} dZ ~e^{-r Z}~f_Z(x_0) \\
&=\langle e^{-r \int _{0}^{t_f} U[x(\tau)] d \tau} \rangle,
\label{def-Q}
\end{align}
where $\langle .. \rangle$ in Eq. \ref{def-Q} represents an average over all possible Brownian paths that start at $x_0$ at $\tau=0$ and stop at their first encounter with the target. To proceed further, consider one realization of a stochastic path $\{ x(\tau);0 \leq \tau \leq t_f \}$ and split that into two parts: (i) the first interval $[0,\Delta t]$ and (ii) the rest interval $[\Delta t,t_f]$ with $\Delta t \to 0$. In the first part, the Brownian path propagates from $x_0$ at $\tau=0$ to $x_0'=x_0 + \Delta x$ in time $\Delta t$. In the rest interval, the process continues until the particle, which is at $x_0'$ at time $\Delta t$, is absorbed at the origin at a later time $t_f$, possibly undergoing many resetting events. This path-decomposition leads us to break the integration in Eq. \eqref{def-Q} as $\int _{0}^{t_f} = \int _{0}^{\Delta t}+\int _{\Delta t}^{t_f}$ and in the limit $\Delta t \to 0$, we get
\begin{align}
\tilde{T}(r,x_0) &= \langle e^{-r U(x_0) \Delta t} ~e^{-r \int _{\Delta t}^{t_f} U[x(\tau)] d \tau} \rangle, \nonumber\\ 
& =\langle e^{-r U(x_0) \Delta t} \tilde{T}(r, x_0+\Delta x) \rangle.
\label{def-Q1}
\end{align}
Here, we have used the fact that, over the second interval $[\Delta t, t_f]$, the initial location itself is a random. The average in the second line of Eq. \ref{def-Q1} is taken over all possible realizations of $\Delta x$. We substitute $\Delta x$ from the underlying process and expanding in the powers of $\Delta t$, and keep only leading-order terms to derive the backward equation.
\begin{align}
\frac{d^2 \tilde{T}(r,x_0)}{d x_0^2} - k_1 x_0\frac{d \tilde{T}(r,x_0)}{d x_0}- r U(x_0) \tilde{T}(r,x_0) = 0. 
\label{bfp}
\end{align}
This equation is valid in the domain $[-\infty,1]$, subject to the following boundary conditions.
\begin{align}
&{\tilde{T}(r, x_0 \to 1) =1,} ~\label{Bbc1} \\
& \tilde{T}(r, x_0 \to -\infty) =0.
\label{Bbc2}
\end{align}
To calculate the Laplace transform of first-passage time distribution function we set $U(x_0)=1$ in the above differential equation and solve with appropriate boundary conditions \ref{Bbc1} and \ref{Bbc2}, we get
\begin{align}
\tilde{T}(r,x_0) = \frac{\pi H\left( -\frac{r}{k_1},\sqrt{\frac{k_1}{2}} x_0 \right)-\Gamma\left(\frac{r}{2k_1}\right)\Gamma\left(1-\frac{r}{k_1}\right)~_1F_1\left(\frac{r}{2k_1},\frac{1}{2},\frac{k_1x_0^2}{2}\right)\sin\left(\frac{\pi r}{k_1}\right) }{\pi H\left( -\frac{r}{k_1},\sqrt{\frac{k_1}{2}}\right)- \Gamma\left(\frac{r}{2k_1}\right)\Gamma\left(1-\frac{r}{k_1}\right)~_1F_1\left(\frac{r}{2k_1},\frac{1}{2},\frac{k_1}{2}\right)\sin\left(\frac{\pi r}{k_1}\right)}
\end{align}
Where $H(n,x)$ is the Hermite polynomial, $\Gamma(z)$ the Euler gamma function and $_1F_1(a,b,z)$ the Kummer confluent hypergeometric function. For the initial location $x_0=0$, the above equation simplifies to
\begin{align}
\tilde{T}(r) = \frac{\frac{2^{-\frac{r}{k_1}}\pi^{\frac{3}{2}}}{\Gamma\left( \frac{1}{2} +\frac{r}{2k_1} \right)} -\Gamma\left(\frac{r}{2k_1}\right)\Gamma\left(1-\frac{r}{k_1}\right)\sin\left(\frac{\pi r}{k_1}\right) }{\pi H\left( -\frac{r}{k_1},\sqrt{\frac{k_1}{2}}\right)- \Gamma\left(\frac{r}{2k_1}\right)\Gamma\left(1-\frac{r}{k_1}\right)~_1F_1\left(\frac{r}{2k_1},\frac{1}{2},\frac{k_1}{2}\right)\sin\left(\frac{\pi r}{k_1}\right)}
    \label{LT_fpt}
\end{align}
\section{Propagator evaluation for underlying process in Laplace space}\label{appC}
We consider the stochastic dynamics governed by the Langevin equation
\begin{align}
   \frac{dx}{dt}= -k_1 x +\sqrt{2}\,\eta(t),
\end{align}
where $\eta(t)$ represents a Gaussian white noise with zero mean and unit
variance, satisfying $\langle \eta(t)\rangle =0$ and 
$\langle \eta(t)\eta(t')\rangle=\delta(t-t')$. The parameter $k_1>0$
characterizes the strength of the restoring force.

The time evolution of the probability density $P(x,t)$ associated with the
stochastic process is governed by the corresponding Fokker--Planck equation
\begin{align}
\frac{\partial P(x,t)}{\partial t} =
k_1\frac{\partial [xP(x,t)]}{\partial x} +
\frac{\partial^2 P(x,t)}{\partial x^2}.
\label{FP}
\end{align}

The process starts from a fixed initial position $x_0$, so that the initial
condition for the probability density is
\begin{align}
    P(x,0)=\delta(x-x_0).
\end{align}
Without loss of generality, we choose the initial position to be $x_0=0$.

The particle evolves in a semi-infinite domain with absorbing boundaries.
Specifically, an absorbing boundary is located at $x=1$, while the left boundary is taken to $x\to -\infty$. Thus the boundary conditions are
\begin{align}
    P(x=1,t)=P(x\to -\infty,t)=0.
\end{align}

To determine the first-passage properties, it is convenient to work in
Laplace space. We therefore define the Laplace transform of the probability
density as
\begin{align}
    \tilde{P}(x,r)=\int_{0}^{\infty} dt\, e^{-rt} P(x,t).
    \label{LT}
\end{align}

Applying the Laplace transform defined in Eq.~(\ref{LT}) to the
Fokker--Planck equation (\ref{FP}) and using the initial condition,
we obtain the following ordinary differential equation for
$\tilde{P}(x,r)$:
\begin{align}
    \frac{d^2\tilde{P}(x,r)}{dx^2}
    +k_1 x\frac{d\tilde{P}(x,r)}{dx}
    +(k_1-r)\tilde{P}(x,r)
    =-\delta(x).
\end{align}

This equation must be solved subject to the Laplace-space boundary
conditions
\begin{align}
    \tilde{P}(x=1,r)=\tilde{P}(x\to -\infty,r)=0.
\end{align}

Solving the above equation with these boundary conditions yields the expression of Laplace transform of the propagator
\begin{align}
    \tilde{P}(x,r) =
    \frac{\begin{aligned}
    &e^{-\frac{k_1x^2}{2}}
    \bigg[
    -x\theta(x)\,{}_1F_1\!\left(\frac{k_1+r}{2k_1},\frac{3}{2},\frac{k_1x^2}{2}\right)
    + {}_1F_1\!\left(\frac{k_1+r}{2k_1},\frac{3}{2},\frac{k_1}{2}\right)
    \\
    &\times
    \left(
    \Gamma\!\left(\frac{r}{2k_1}\right)
    {}_1F_1\!\left(\frac{r}{2k_1},\frac{1}{2},\frac{k_1x^2}{2}\right)
    +\sqrt{2k_1}\,x
    \Gamma\!\left(\frac{k_1+r}{2k_1}\right)
    {}_1F_1\!\left(\frac{k_1+r}{2k_1},\frac{3}{2},\frac{k_1x^2}{2}\right)
    \right)
    \bigg]
    \end{aligned}}
    {\Gamma\!\left(\frac{r}{2k_1}\right)
    {}_1F_1\!\left(\frac{r}{2k_1},\frac{1}{2},\frac{k_1}{2}\right)
    +\sqrt{2k_1}\Gamma\!\left(\frac{k_1+r}{2k_1}\right)
    {}_1F_1\!\left(\frac{k_1+r}{2k_1},\frac{3}{2},\frac{k_1}{2}\right)}.\label{LT_prop}
\end{align}
Here ${}_1F_1(a,b,z)$ denotes the Kummer confluent hypergeometric function,
$\theta(x)$ is the Heaviside step function defined as $\theta(x)=0$ for
$x<0$ and $\theta(x)=1$ for $x>0$, and $\Gamma(z)$ represents the Euler
gamma function.
\section{First passage statistics of diffusing particle during the return phase}\label{appD}
The Fokker–Planck equation associated with the Langevin dynamics in the return phase takes the form
\begin{align}
    \frac{\partial P_{\text{ret}}(x,t)}{\partial t} = k_2\frac{\partial [xP_{\text{ret}}(x,t)]}{\partial x} + \frac{\partial^2 P_{\text{ret}}(x,t)}{\partial x^2}
    \label{FP_ret}
\end{align}
with the initial condition; 
\begin{align}
    P(x,t=0)=\delta(x-x_0)
\end{align}
where $x_0$ is the initial location of the searcher at the time of resetting. The boundary condition is given by
\begin{align}
    P_{\text{ret}}(x=0,t) = P_{\text{ret}}(x=1,t) = 0
\end{align}
Writing above Eq. \ref{FP_ret} in Laplace space,
\begin{align}
    \frac{d^2\tilde{P}_{\text{ret}}(x,r)}{dx^2} +k_2x\frac{d\tilde{P}_{\text{ret}}(x,r)}{dx} + (k_2-r)\tilde{P}_{\text{ret}}(x,r) = -\delta(x-x_0)
    \label{BFP_ret}
\end{align}
Since we are interested in the time-integrated quantity during the return phase, we instead solve a simplified differential equation by setting $r = 0$ in the above Eq. \ref{BFP_ret}.
\begin{align}
    \frac{d^2\tilde{P}_{\text{ret}}(x,0)}{dx^2} +k_2x\frac{d\tilde{P}_{\text{ret}}(x,0)}{dx} + k_2\tilde{P}_{\text{ret}}(x,0) = -\delta(x-x_0)
\end{align}
\begin{align}
    \tilde{P}_{\text{ret}}(x)=\begin{aligned}
        &\left(\sqrt{k_2}\text{Erfi}\Big(\sqrt{\frac{k_2}{2}}\Big)\right )^{-1}e^{-k_2x^2/2}\sqrt{\frac{\pi}{2}}\pmb{\bigg[ } \text{Erfi}\Big(\sqrt{\frac{k_2}{2}}x\Big)\Big[\text{Erfi}\Big(\sqrt{\frac{k_2}{2}}\Big) - \text{Erfi}\Big(\sqrt{\frac{k_2}{2}}x_0\Big)\Big]\\
        &+\text{Erfi}\Big(\sqrt{\frac{k_2}{2}}\Big) \Big[ \text{Erfi}\Big(\sqrt{\frac{k_2}{2}}x_0\Big) - \text{Erfi}\Big(\sqrt{\frac{k_2}{2}}x\Big) \Big ]\theta(x - x_0)\pmb{\bigg ]}
    \end{aligned}
\end{align}
where $\text{Erfi}(z)=\text{erf}(iZ)/i$ is the imaginary error function and $i=\sqrt{-1}$ and $\text{erf}(z)$ is the error function.\newline
\textbf{The Mean First passage Time}
The survival probability $Q(t)$ of the particle is given as
\begin{align}
    Q(t) = \int_{0}^{1} dx~ P_{\text{ret}}(x,t)
\end{align}
which in Laplace space becomes
\begin{align}
    \tilde{Q}(s) = \int_{0}^{\infty} dt~e^{-st}Q(t) = \int_{0}^{1} dx~ \tilde{P}_{\text{ret}}(x,s)
\end{align}
The unconditional MFPT for the particle to exit through either boundary is given by
\begin{align*}
    \langle t_2(x_0) \rangle = \lim_{s \to 0}\tilde{Q}(s) 
\end{align*}
and given by
\begin{align}
    \langle t_2(x_0) \rangle =~ \begin{aligned}
        & \frac{1}{2} \pmb{\Bigg [ } \frac{\pi}{k_2} \text{Erfi}\Big(\sqrt{\frac{k_2}{2}}x_0\Big) \Big[ \text{Erf}\Big(\sqrt{\frac{k_2}{2}}\Big)-\text{Erf}\Big(\sqrt{\frac{k_2}{2}}x_0\Big) \Big] + x_0^2~ _2F_2\Big[(1,1),\Big
        (\frac{3}{2},2\Big),-\frac{k_2x_0^2}{2}\Big] \\
        & - \frac{\text{Erfi}\Big(\sqrt{\frac{k_2}{2}}x_0\Big)~_2F_2\Big[(1,1),(\frac{3}{2},2),-\frac{k_2}{2}\Big]}{\text{Erfi}\Big(\sqrt{\frac{k_2}{2}}\Big)}  \pmb{\Bigg ]}\label{mfpt2}
    \end{aligned}
\end{align}
where, $_pF_q[(a_1,...,a_p),(b_1,...,b_q),z]$ is the generalized hypergeometric function.\newline
Now, when the particle resides in the negative side of the quadrant, it encounters only a single boundary at the origin, and the corresponding mean first-passage time is given by
\begin{align}
    \langle t_1(x_0) \rangle = \frac{1}{2}\pmb{\bigg[} \frac{\pi}{k_2}\text{Erfi}\Big(\sqrt{\frac{k_2}{2}}x_0\Big)\Big[ 1 - \text{Erfi}\Big(\sqrt{\frac{k_2}{2}}x_0\Big) \Big] +  x_0^2~ _2F_2\Big[(1,1),\Big
        (\frac{3}{2},2\Big),-\frac{k_2x_0^2}{2}\Big]\pmb{\bigg ]}\label{mfpt1}
\end{align}
\textbf{Splitting Probability}\newline
The probability flux through each of the boundary in the Laplace space is given by
\begin{align}
    &j_1(x_0,r) = -D\frac{\partial\tilde{P}_{\text{ret}}(x,r)}{\partial x}\Bigg |_{x=L}\\
    &j_0(x_0,r) = ~D\frac{\partial\tilde{P}_{\text{ret}}(x,r)}{\partial x}\Bigg |_{x=0}
\end{align}
\textbf{Splitting/Exit probabilities}\newline
The corresponding exit probabilities are given by the
\begin{align}
    &\epsilon_1(x_0) = j_1(x_0, r\to 0)= \frac{\text{Erfi}\Big(\sqrt{\frac{k_2}{2}}x_0\Big)}{\text{Erfi}\Big(\sqrt{\frac{k_2}{2}}\Big)}\label{splitting_target}\\
    &\epsilon_0(x_0) = j_0(x_0, r\to 0)= 1-\frac{\text{Erfi}\Big(\sqrt{\frac{k_2}{2}}x_0\Big)}{\text{Erfi}\Big(\sqrt{\frac{k_2}{2}}\Big)}\label{splitting_origin}
\end{align}

\bibliography{SwitchingTrap}

\end{document}